\documentclass[aps,pra,twocolumn,showpacs,superscriptaddress]{revtex4}

\usepackage{amsfonts}
\usepackage{amsmath}
\usepackage{amssymb}
\usepackage{epsfig}
\usepackage{graphicx}

\begin{document}

\title{Strongly-Driven One-Atom Laser and Decoherence Monitoring}

\date{\today}

\author{P. Lougovski}
\affiliation{Hearne Institute for Theoretical Physics, Department of
Physics and Astronomy, Louisiana State University, 202 Nicholson
Hall, Baton Rouge, LA 70803, USA}

\author{F. Casagrande}
\affiliation{Dipartimento di Fisica, Universit\`{a} di Milano, Via
Celoria 16, 20133 Milano, Italy }

\author{A. Lulli}
\affiliation{Dipartimento di Fisica, Universit\`{a} di
Milano, Via Celoria 16, 20133 Milano, Italy }

\author{E. Solano}
\affiliation{Physics Department, ASC, and CeNS,
Ludwig-Maximilians-Universit\"at, Theresienstrasse 37, 80333 Munich,
Germany} \affiliation{Secci\'on F\'{\i}sica, Departamento de
Ciencias, Pontificia Universidad Cat\'olica del Per\'u, Apartado
1761, Lima, Peru}


\begin{abstract}
We propose the implementation of a strongly-driven one-atom laser,
based on the off-resonant interaction of a three-level atom in
$\Lambda$-configuration with a single cavity mode and three laser
fields. We show that the system can be described equivalently by a
two-level atom resonantly coupled to the cavity and driven by a
strong effective coherent field. The effective dynamics can be
solved exactly, including a thermal field bath, allowing an
analytical description of field statistics and entanglement
properties. We also show the possible generation of Schr\"odinger
cat states for the whole atom-field system and for the field alone
after atomic measurement. We propose a way to monitor the system
decoherence by measuring atomic population. Finally, we confirm
the validity of our model through numerical solutions.
\end{abstract}
\maketitle
\section{Introduction}
In cavity quantum electrodynamics (CQED) the interaction between
atoms and photons can be investigated experimentally under carefully
controlled conditions, and described by relatively simple
models~\cite{intro1}. These features make CQED an almost ideal
framework to investigate the foundations of quantum mechanics and
their application to quantum information~\cite{intro2}. For
instance, two-atom entanglement~\cite{intro3} as well as the
entanglement between an atom and a photon~\cite{intro4} have been
recently demonstrated. On the other hand, the basic interaction
between a two-level atom and a cavity field mode, as described by
the Jaynes-Cummings (JC) model~\cite{intro5}, leads to nonclassical
effects carefully tested in recent years~\cite{intro6}. Furthermore,
it allowed the implementation of the micromaser~\cite{intro7} and
the microlaser~\cite{intro8} in the strong coupling regime of CQED,
in the microwave and optical domain, respectively. Further efforts
led to the implementation of a trapped ion as a nanoscopic probe of
cavity field modes~\cite{intro9}. More recently, a single trapped
neutral atom in a high-Q optical cavity~\cite{intro10} allowed the
implementation of a one-atom laser~\cite{intro11}, i.e., lasing with
only one intra-cavity atom. These systems can exhibit features that
are not present in standard macroscopic lasers such as thresholdless
generation and sub-poissonian photon number distribution~\cite{intro12}.\\
\indent Another milestone in CQED experiments was reached in
Ref.~\cite{intro13}, where a ``Schr\"{o}dinger cat'' state of the
cavity field, a mesoscopic superposition of two coherent states, was
realized. There, the field decoherence was monitored through
atom-atom correlation measurements~\cite{Davidovich}. State
reconstruction of nonclassical intra-cavity fields was also possible
through atom-cavity dispersive interactions~\cite{intro14,intro15}.
More recently, a remarkable proposal for the resonant generation of
Schr\"odinger cat states~\cite{intro16} was implemented in the
lab~\cite{intro17}, and tested with the help of a quantum spin-echo
technique~\cite{intro18}. The understanding of entanglement in
atom-cavity systems was enhanced when an additional driving field
acting on the cavity mode was added on top of the atom-cavity JC
interaction~\cite{intro19,intro20}. In this respect, recently, an
elegant analysis of a driven cavity containing a two-level atom
explained the absence or increase of entanglement in the transient
of the atom-cavity dynamics~\cite{intro21}. Unfortunately, most
realistic models including dissipative processes require numerical
analysis, or ideal theoretical conditions for the sake of
semi-analytical derivations.\\
\indent In this work, we introduce an integrable model of a
strongly-driven one-atom laser (SDOAL) operating in the optical
regime of CQED, where the coherent driving field acts directly on
the atom. We consider a realistic model consisting of a three-level
atom in $\Lambda$-configuration placed inside a single-mode optical
cavity, coupled off-resonantly to three coherent laser fields. We
show that this model can be reduced to two atomic levels coupled to
a cavity mode and a strong classical driving on the atom. In this
strong-driving limit~\cite{intro22}, we solve the full system
dynamics~\cite{intro23}, in the transient and in the steady state,
providing one more of the few examples of an exactly solvable open
quantum system.  In previous works~\cite{intro24,intro25}, we
developed related results for microwave cavity fields and two-level
Rydberg atoms, not a good model for a field in the optical regime
and fast decaying atomic dipolar transitions. Here, we solve
analytically the master equation for the full atom-field system in
the SDOAL model. Next, we exploit the obtained solutions for the
analysis of atom-field entanglement and the decoherence of
atom-field superposition states ( Schr\"odinger cat states) via the
measurement of atomic populations. In addition, the generation and
the decoherence of cat states of the cavity field alone is
described. Finally, we present numerical results confirming the
validity of the approximations made to derive
the effective master equation of the SDOAL model.\\
\indent The paper is organized as follows. In section II we
introduce the integrable model of a SDOAL. In section III we solve
analytically the master equation for the atom-cavity dynamics. In
section IV we consider the dynamics of the cavity field and atom
subsystems. In section V we describe entanglement properties and the
environment-induced decoherence of the SDOAL, presenting a scheme to
monitor decoherence via atomic populations measurements. A numerical
analysis that confirms the validity of the model is presented in
section VI. Conclusions are reported in section VII.

\section{The strongly-driven one-atom laser model}
\begin{figure}[h]
\begin{center}
\includegraphics[scale=0.40]{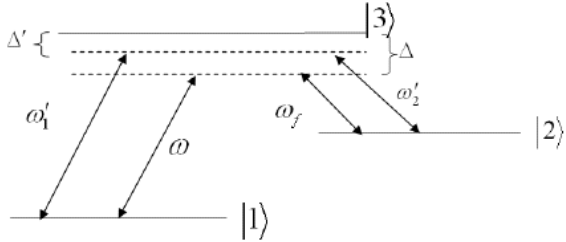}\vspace*{-0.5cm}
\end{center}
\caption{Atomic energy levels and the applied fields. $\Delta$ and
$\Delta^{\prime}$ denote the frequency detunings, $\omega_f$ is the
frequency of the cavity mode, and $\omega_{1}^{\prime}$,
$\omega_{2}^{\prime}$, $\omega$ are the frequencies of the lasers
applied to the associated transitions.} \vspace*{-0.5cm}
\label{fig:lambdascheme}
\end{figure}

We consider a three-level atom (ion) in a $\Lambda$-configuration
trapped inside an optical cavity (Fig.~\ref{fig:lambdascheme}). We
assume that the transition $|1\rangle\leftrightarrow|2\rangle$ is
quadrupolar and, hence, the metastable states $|1\rangle$ and
$|2\rangle$ cannot be coupled directly, but only via the level
$|3\rangle$. The level $|3\rangle$ can decay via spontaneous
emission and, therefore, the external lasers and the cavity field
are all far detuned with respect to the corresponding transition
frequencies. We suppose that the atom interacts off-resonantly with
a single mode of a cavity field of a frequency $\omega_f$ on the
transition $|3\rangle\leftrightarrow|2\rangle$. The same transition
is coupled off-resonantly to a coherent field of a frequency
$\omega_{2}^{\prime}$. The remaining atomic transition
$|3\rangle\leftrightarrow|1\rangle$ is coupled off resonance to two
lasers of frequency $\omega_{1}^{\prime}$ and $\omega$. The
different frequency detunings, $\Delta$ and
$\Delta^{\prime}<\Delta$, of these two $\Lambda$-processes prevent
the system from undesired transitions.
\\
\\
We assume, without loss of generality, that both the
cavity mode coupling frequency $g$ and the associated laser Rabi
frequencies $\Omega,\Omega'_1, \Omega'_2 $ are real. The Hamiltonian
$\hat{\mathcal{H}}(t)$ for the whole system can be written as
$\hat{\mathcal{H}}(t)=\hat{\mathcal{H}_0}+\hat{\mathcal{H}_1}(t)$,
where
\begin{equation}
\label{eq:hamLambda} \hat{\mathcal{H}_0}  =  \hbar\omega_{3}
\hat{S}^{33} +
\hbar\omega_{2}\hat{S}^{22}+\hbar\omega_{1}\hat{S}^{11} +
\hbar\omega_{f}\hat{a}^{\dagger}\hat{a},
\end{equation}
\begin{eqnarray}
\label{eq:hamLambda1} \hat{\mathcal{H}}_1(t) & = &\hbar g
(\hat{a}^{\dagger}\hat{S}^{23}_{-} +
\hat{a}\hat{S}^{23}_{+})+\hbar\Omega (e^{-i\omega
t}\hat{S}^{13}_{+}+e^{i\omega t}\hat{S}^{13}_{-}) \nonumber\\
& + & \hbar \Omega'_1(e^{-i\omega'_{1}t}\hat{S}^{13}_{+} +
e^{i\omega'_{1}t}\hat{S}^{13}_{-}) \nonumber\\  & + &
\hbar\Omega'_2(e^{-i\omega'_{2}t}\hat{S}^{23}_{+}+e^{i\omega'_{2}t}
\hat{S}^{23}_{-}).
\end{eqnarray}
Here $\hat{a}$ ($\hat{a}^{\dagger}$) is the cavity mode annihilation
(creation) operator and, following the notation of~\cite{Klimov}, we
define the atomic operators as follows,
\begin{eqnarray}
\label{atomicoperators} \hat{S}^{23}_{+} &=&
|3\rangle\langle2|,\hat{S}^{23}_{-}=|2\rangle\langle
3|,\hat{S}^{13}_{+}=|3\rangle\langle 1|,\nonumber \\
\hat{S}^{13}_{-} &=& |1\rangle\langle
3|,\hat{S}^{JJ}=|J\rangle\langle J|,\hspace{0.5cm}(J=1,2,3).
\end{eqnarray}
We rewrite the Hamiltonian $\hat{\mathcal{H}}(t)$ in the
interaction picture leaving the unavoidable time dependence in the
term related to the laser Rabi frequency $\Omega$
\begin{eqnarray}
\label{H_interaction} \hat{\mathcal{H}_i}(t) & = &
-\hbar\Delta'{\hat{S}}^{22}-\hbar\Delta'{\hat{S}}^{11} -
\hbar(\Delta-\Delta')\hat{a}^{\dagger}\hat{a} \nonumber \\ & + &
\hbar g [(\hat{a}+\frac{\Omega'_2}{g})\hat{S}^{23}_{+}+h.c.]
\nonumber \\ & + &
\hbar\Omega'_1[(1+\frac{\Omega}{\Omega'_1}e^{i(\Delta-\Delta')t})
\hat{S}^{13}_{+} + {\rm H.c.}].
\end{eqnarray}

If $\{ \Delta , \Delta' , | \Delta - \Delta'| \} \gg \{g , \Omega ,
\Omega_1' , \Omega_2' \}$, the unitary dynamics of the atom-field
system in Eq.~(\ref{H_interaction}) can be described by an effective
Hamiltonian for a two-level atom coupled to the cavity mode and in
presence of a classical driving field. This is due to the fact that,
under these conditions, the energy diagram of
Fig.~\ref{fig:lambdascheme} can be understood as composed by two
independent $\Lambda$-schemes. In this case, it is straightforward
to prove that we can build the second-order Hamiltonian
\begin{eqnarray}
\label{effectiveHamiltonian1} \hat{\bar{\mathcal{H}'}}_{\rm eff}  = -  \hbar \bar{g}_{\rm
eff}(\hat{a}^{\dagger}\hat{S}^{12}_{+}+\hat{a}\hat{S}^{12}_{-}) -
\hbar \bar{\Omega}_{\rm eff}(\hat{S}^{12}_{+}+\hat{S}^{12}_{-}) ,
\end{eqnarray}
with $\bar{g}_{\rm eff} = g \Omega / \Delta$ and
$\bar{\Omega}_{\rm eff} = \Omega'_1 \Omega_2' / \Delta'$. In
Eq.~(\ref{effectiveHamiltonian1}), as is usually done, we have
assumed the compensation of constant AC Stark-shift terms by a
proper retuning of the laser frequencies. The Stark-shift term
depending on the intracavity photon number can be neglected if
$\Omega \gg g$. In the strong-driving limit, $\bar{\Omega}_{\rm
eff} \gg \bar{g}_{\rm eff}$, as explained in Ref.~\cite{intro22},
we can derive the final effective Hamiltonian
\begin{eqnarray}
\label{eq:effectiveHamiltonian2}\hat{\bar{\mathcal{H}}}_{\rm eff} & = & - \hbar \frac{
\bar{g}_{\rm eff}} {2} (\hat{a}^{\dagger} + \hat{a})(\hat{S}^{12}_{+}
+\hat{S}^{12}_{-}) .
\end{eqnarray}
We have tested numerically the above analytical considerations and
proved, in fact, that Eq.~(\ref{eq:effectiveHamiltonian2})
describes the correct effective dynamics. However, we want to show
here that we can go beyond the limit of uncoupled
$\Lambda$-schemes and obtain a similar dynamics with less
demanding conditions on the experimental parameters. To prove this
statement we exploit the small rotations method of
Ref.~\cite{Klimov}, which is essentially a perturbative method for
deriving effective Hamiltonians. First, we introduce the operators
of a $SU(3)$ deformed algebra,
\begin{eqnarray}
\label{su3operators} \hat{X}^{23}_{+} &=&
(\hat{a}+\frac{\Omega'_2}{g})\hat{S}^{23}_{+} \,\, ,
\hspace{0.5cm}\hat{X}^{23}_{-} = (\hat{a}^{\dagger} +
\frac{\Omega'_2}{g})\hat{S}^{23}_{-} , \nonumber\\
\hat{Y}^{13}_{+} &=&b_t \hat{S}^{13}_{+} \,\, , \hspace{0.5cm}
\hat{Y}^{13}_{-} =b_t^{*} \hat{S}^{13}_{-} ,
\end{eqnarray}
where $b_t=1+\frac{\Omega}{\Omega'_1}e^{i(\Delta-\Delta')t}$. Using
the identity relation $\hat{I}=\hat S^{11}+\hat S^{22}+\hat S^{33}$
we can rewrite the interaction Hamiltonian Eq.~(\ref{H_interaction})
in the compact form,
\begin{eqnarray}
\label{H_interactionRWA} \hat{\mathcal{H}_i}(t) & = &
-\hbar\Delta'+\hbar\Delta'\hat{S}^{33}- \hbar(\Delta-\Delta')
\hat{a}^{\dagger}\hat{a} \nonumber \\ & + &  \hbar
g(\hat{X}^{23}_{+}+\hat{X}^{23}_{-})
+\hbar\Omega'_1(\hat{Y}^{13}_{-}+\hat{Y}^{13}_{-}).
\end{eqnarray}
We can eliminate the dependence of $\hat{\mathcal{H}_i}$ on the
upper level $|3\rangle$ by applying two consecutive small rotations.
The first unitary transformation
$\hat{U}^{13}=\exp\{\alpha(\hat{Y}^{13}_{+}-\hat{Y}^{13}_{-}) \}$,
with $\alpha=\frac{\Omega'_1}{\Delta'}\ll1$ and the condition
$(\frac{\Omega}{\Delta'})^2\ll 1$, allows us to eliminate the
dependence on operators $\hat{Y}^{13}_{\pm}$. The second unitary
small rotation, given by
$\hat{U}^{23}=\exp\{\beta(\hat{X}^{23}_{+}-\hat{X}^{23}_{-}) \}$,
with $\beta=\frac{g}{\Delta}\ll1$ and the conditions
$(\frac{\Omega'_2}{\Delta})^2\ll 1$,
$(\frac{\Delta-\Delta'}{\Delta})^2\ll 1$, can be used to eliminate
the dependence on $\hat{X}^{23}_{\pm}$. After some lengthy algebra
we derive the effective two-level Hamiltonian
\begin{eqnarray}
\!\!\! \label{H_effectivetilda} \hat{\mathcal{H}'}_{\rm eff} & = &
\hbar(\Delta-\Delta')\hat{S}^{22}+\hbar(\Delta-\Delta')
\hat{a}^{\dagger}\hat{a} \nonumber \\ & - & \hbar g_{\rm
eff}(\hat{a}^{\dagger}\hat{S}^{12}_{+}+\hat{a}\hat{S}^{12}_{-}) -
\hbar\Omega_{\rm eff}(\hat{S}^{12}_{+}+\hat{S}^{12}_{-}) ,
\end{eqnarray}
where we introduced the effective coupling and driving frequencies
$g_{\rm eff}=\frac{g\Omega}{\Delta'}$ and $\Omega_{\rm
eff}=\frac{\Omega\Omega'_2}{\Delta'}$. Note that the above
derivation does not depend on the order of the two small rotations.
The effective Hamiltonian (\ref{H_effectivetilda}) is exact to
zero-th order in the diagonal terms and to first-order in the other
ones. From now on we shall consider the case of small detuning
difference $\frac{\Delta-\Delta'}{\Delta}\ll 1$, such that the
diagonal terms are negligible. This will allow us to obtain an
exactly solvable model of system dynamics even in the presence of
dissipation, that is a one-atom laser. Hence, the initial model
described by the Hamiltonian of Eqs. (\ref{eq:hamLambda}) and
(\ref{eq:hamLambda1}) reduces to a Hamiltonian that exhibits an
effective coupling of the states $|1\rangle$ and $|2\rangle$ to the
cavity mode in the presence of a classical external field driving
the atomic transition. Now we can apply the unitary transformation
$\hat{U} = \exp\{-i\Omega_{\rm eff}(\hat{S}^{12}_{+} +
\hat{S}^{12}_{-})t\}$ to obtain~\cite{intro22}
\begin{eqnarray}
\label{Hint4} \hat{\mathcal{H}''}_{\rm eff} & = & -\hbar\frac{g_{\rm
eff}}{2}[ (|+\rangle\langle+|-|-\rangle\langle-| + e^{-2i\Omega_{\rm
eff} t}|+ \rangle\langle-| \nonumber \\ & - &  e^{2i\Omega_{\rm eff}
t}|-\rangle\langle+|)\hat{a}^{\dagger} + {\rm H.c.}],
\end{eqnarray}
where we used the eigenstates
$|\pm\rangle=\frac{|1\rangle\pm|2\rangle}{\sqrt{2}}$ of the operator
$\hat{S}_{x}\equiv\hat{S}^{12}_{+} + \hat{S}^{12}_{-}$. In this way,
we put in evidence fast rotating terms in Eq.~(\ref{Hint4}) and, after
applying the RWA with $\Omega_{\rm eff} \gg g_{\rm eff}$, we obtain the final effective
Hamiltonian
\begin{eqnarray}
\label{eq:keyLas}\hat{\mathcal{H}}_{\rm eff} & = & - \hbar \frac{
g_{\rm eff}} {2} (\hat{a}^{\dagger} + \hat{a})(\hat{S}^{12}_{+}
+\hat{S}^{12}_{-}).
\end{eqnarray}
This Hamiltonian has the structure of resonant and simultaneous
Jaynes-Cummings and 'anti-Jaynes-Cummings'
interactions~\cite{intro22} and its dynamics is better understood in
terms of Schr\"odinger cat states than Rabi oscillations, as will be
discussed later. Note that the Hamiltonian of Eq.~(\ref{eq:keyLas})
is similar to the one of Eq.~(\ref{eq:effectiveHamiltonian2}) but
with a more relaxed set of parameters. Furthermore, the dynamics is
fully confirmed by
numerical simulations.\\
\indent To describe the open atom-cavity system dynamics we must
include the dissipative effects due to the coupling of the cavity to
the environment. We note that the decay of the upper level
$|3\rangle$ can be neglected because of the elimination procedure
described above. Therefore, the system dynamics can be described by
the following SDOAL master equation (ME)
\begin{eqnarray}
\label{eq:maseqL}\dot{\rho}_{AF} & = & -
\frac{i}{\hbar}[\hat{\mathcal{H}}_{\rm eff}, \rho_{AF}] +
\hat{\mathcal{L}}\rho_{AF},
\end{eqnarray}
where the dissipative term is the standard Liouville superoperator
for a damped harmonic oscillator
\begin{eqnarray}
\label{eq:dissip}\hat{\mathcal{L}}\rho_{AF} & = &
-\frac{\kappa}{2}(\hat{a}^{\dagger}\hat{a}\rho_{AF} -
2\hat{a}\rho_{AF}\hat{a}^{\dagger} +
\rho_{AF}\hat{a}^{\dagger}\hat{a}) .
\end{eqnarray}
Here, $\kappa$ is the cavity photon decay rate and we consider the
limit of zero temperature because the system operates in the optical
regime.

\section{ANALYTICAL SOLUTION OF THE SDOAL MASTER EQUATION}

The time evolution of the atom-field system is described by the
density operator $\rho_{AF}(t)$ which is the solution of the ME in
Eq.~(\ref{eq:maseqL}). In order to solve it, we introduce the
following decomposition
\begin{eqnarray}\label{eq:lasatfdenop}
\rho_{AF}(t) & = & |+\rangle\langle +|\otimes\rho_{1F}(t) +
|-\rangle\langle -|\otimes\rho_{2F}(t) \nonumber \\ & + &
|+\rangle\langle -|\otimes\rho_{3F}(t) + |-\rangle\langle
+|\otimes\rho_{4F}(t)
\end{eqnarray}
Here, $\rho_{iF}(t)$ ($i = 1,...,4$ ) are operators describing the
cavity field defined as
\begin{eqnarray}\label{eq:roifdef}
\rho_{1F}(t) &  =  & \langle + |\rho_{AF}(t)| + \rangle,
\hspace{0.2cm}\rho_{2F}(t) = \langle-|\rho_{AF}(t)| - \rangle,
\nonumber
\\ \rho_{3F}(t) & = &
\langle+|\rho_{AF}(t)|-\rangle,\hspace{0.2cm}\rho_{4F}(t) =
\langle-|\rho_{AF}(t)|+\rangle . \nonumber \\
\end{eqnarray}
Then, the master equation (\ref{eq:maseqL}) is equivalent to the
following set of equations for the operators $\rho_{iF}(t)$
\begin{eqnarray}
\label{eq:equivsys1}\dot{\rho}_{1,2F} & = & \pm i \frac{ g_{\rm
eff}}{2}[\hat{a}^{\dagger}+\hat{a},\rho_{1,2F}] +
\hat{\mathcal{L}}\rho_{1,2F}, \\
\label{eq:equivsys2} \dot{\rho}_{3,4F} & = & \pm i\frac{g_{\rm
eff}}{2}\{\hat{a}^{\dagger}+\hat{a},\rho_{3,4F}\} +
\hat{\mathcal{L}}\rho_{3,4F} ,
\end{eqnarray}
where brackets $\lbrack \, , \rbrack$ and braces $\lbrace \, ,
\rbrace$ denote the standard commutator and anti-commutator symbols.
In order to describe a one-atom laser  dynamics we assume that the
initial atom-field density operator is
$\rho_{AF}(0)=|1\rangle\langle1|\otimes|0\rangle\langle0|$.
Therefore, operators $\rho_{iF}(0)$ read
\begin{equation}\label{eq:incondfield}
\rho_{iF}(0) = \frac{1}{2}|0\rangle\langle 0|\hspace{1cm}
(i=1,...,4).
\end{equation}
This choice is suitable in the optical regime of CQED also because
the generation of coherent states is difficult due to the very fast
decay of the cavity mode. Nevertheless, from a theoretical point of
view and for an extension to the microwave regime of CQED it is
possible to generalize the following analysis to the case of a field
prepared in a coherent state. The results in Eqs.~(\ref{roiFt})
should be modified by redefining the form of the
function $\alpha(t)$. \\
In order to solve Eqs.~(\ref{eq:equivsys1})
and (\ref{eq:equivsys2}), we map them onto a set of first order
partial differential equations for the functions
$\chi_i(\beta,t)=\textrm{Tr}_F\lbrack \rho_{iF}(t)\hat{D}(\beta)
\rbrack$, $i=1,...,4$, where $\hat{D}(\beta)$ denotes a displacement
operator. The functions $\chi_{i}(\beta, t)$ cannot be interpreted
as characteristic functions for the cavity field, because the
operators $\rho_{iF}(t)$ do not exhibit all required properties of a
density operator. As a consequence the functions $\chi_{i}(\beta,
t)$ do not fulfill all conditions for quantum characteristic
functions. Nevertheless, they are continuous and square-integrable,
which is enough for our purposes. From Eqs.~(\ref{eq:equivsys1}) and
(\ref{eq:equivsys2}) we obtain the following set of partial
differential equations,
\begin{eqnarray}\label{eq:cheqsys1}
\frac{\partial\chi_{1,2}}{\partial t} &=& \mp i \frac{ g_{\rm
eff}}{2}(\beta + \beta^{\ast})\chi_{1,2} -
\frac{\kappa}{2}|\beta|^{2}\chi_{1,2} , \nonumber \\ & - &
\frac{\kappa}{2}(\beta\frac{\partial}{\partial\beta} +
\beta^{\ast}\frac{\partial}{\partial\beta^{\ast}})\chi_{1,2}
\end{eqnarray}
\begin{eqnarray}
\label{eq:cheqsys2} \frac{\partial\chi_{3,4}}{\partial t} &=& \pm i
g_{\rm eff}(\frac{\partial}{\partial\beta} -
\frac{\partial}{\partial\beta^{\ast}})\chi_{3,4} -
\frac{\kappa}{2}|\beta|^{2}\chi_{3,4}\nonumber \\ & - &
\frac{\kappa}{2}(\beta\frac{\partial}{\partial\beta} +
\beta^{\ast}\frac{\partial}{\partial\beta^{\ast}})\chi_{3,4}.
\end{eqnarray}
To solve these differential equations we use the method of
characteristics, for which it is useful to rewrite them in terms of
the real and imaginary parts of the complex variable $\beta=x+iy$,
\begin{eqnarray}\label{eq:cheqsys1_xy}
\frac{\partial\chi_{1,2}}{\partial t} \!\! & = & \!\! \mp i g_{\rm
eff}x \chi_{1,2} \nonumber \\ & - & \!\!
\frac{\kappa}{2}(x^{2}+y^{2})\chi_{1,2}
-\frac{\kappa}{2}(x\frac{\partial}{\partial x } + y
\frac{\partial}{\partial y})\chi_{1,2} \, ,
\end{eqnarray}
\begin{eqnarray}
\label{eq:cheqsys2_xy} \frac{\partial\chi_{3,4}}{\partial t} & = &
\pm g_{\rm eff}\frac{\partial\chi_{3,4}}{\partial y}  -
\frac{\kappa}{2}(x^{2}+y^{2})\chi_{3,4} \nonumber \\ & - &\!\!
\frac{\kappa}{2}(x\frac{\partial}{\partial x} + y
\frac{\partial}{\partial y})\chi_{3,4}.
\end{eqnarray}
If in the equations for $\chi_{3,4}$ we introduce the shifted
variable $\tilde{y}=y\mp 2\frac{g_{\rm eff}}{k}$ the above equations
can be written as
\begin{eqnarray}\label{eq:cheqsys1_xy_H12}
\frac{\partial\chi_{1,2}}{\partial
t}+\frac{\kappa}{2}(x\frac{\partial}{\partial x } + y
\frac{\partial}{\partial y})\chi_{1,2} &=& H_{1,2}(x,y)\chi_{1,2}
, \\
\label{eq:cheqsys1_xy_H34}\frac{\partial\chi_{3,4}}{\partial
t}+\frac{\kappa}{2}(x\frac{\partial}{\partial x } + \tilde{y}
\frac{\partial}{\partial \tilde{y}})\chi_{3,4} &=&
H_{3,4}(x,\tilde{y})\chi_{3,4}
\end{eqnarray}
where $H_{1,2}(x,y)=x[F'_1(x)\mp F'_2(x)]+yG'(y)$ and
$H_{3,4}(x,\tilde{y})=\tilde{y}[E'_1(\tilde{y})\mp
E'_2(\tilde{y})]+xD'(x)$. There, we have also introduced the
derivatives of the following functions
\begin{eqnarray}\label{eq:F12G}
F_1(x) & = & -\frac{\kappa}{4}x^2, \hspace{0.2cm}F_2(x) = ig_{\rm
eff}x,
\hspace{0.2cm}G(y) = -\frac{\kappa}{4}y^2,\nonumber\\
E_1(\tilde{y}) & = & -\frac{\kappa}{4}\tilde{y}^2,
\hspace{0.2cm}E_2(\tilde{y}) = 2g_{\rm eff}(\frac{g_{\rm
eff}}{\kappa} \ln\tilde{y}+\tilde{y}), \nonumber\\ D(x) &=&
-\frac{\kappa}{4}x^2.
\end{eqnarray}
With these definitions, together with the initial functions
$\chi_{i,0}(x,y)=\chi_{i}(x,y,0)$ associated with the ones in
Eq.~(\ref{eq:incondfield}), we can write the time-dependent
solutions as
\begin{eqnarray}\label{eq:explsolChi12}
\chi_{1,2}(x,y,t) = \frac{1}{2} \exp \{ -\frac{x^{2} + y^{2}}{2} \mp
2i \frac{g_{\rm eff}x}{\kappa}[1-e^{-\frac{\kappa t}{2}}] \} ,
\nonumber \\
\end{eqnarray}
\begin{eqnarray}\label{eq:explsolChi34}
\chi_{3,4}(x,y,t) = \frac{f(t)}{2} \exp \{-\frac{x^{2} + y^{2}}{2}
\mp 2 \frac{ g_{\rm eff} y}{\kappa}[1-e^{-\frac{\kappa t} {2}}] \} ,
\nonumber \\
\end{eqnarray}
where
\begin{equation}\label{eq:ft}
f(t)  =\exp\left\{-2 \frac{g_{\rm eff}^2}{\kappa}t+ 4 \frac{g_{\rm
eff}^2}{\kappa^2}(1 - e^{-\kappa t/2})\right\} .
\end{equation}
The most striking feature of the solutions for $\chi_{3,4}$ in
Eq.~(\ref{eq:explsolChi34}) is the presence of the factor
$e^{-\frac{2 g_{\rm eff}^{2}}{\kappa}t}$, in contrast to the
solutions for $\chi_{1,2}$ in Eq.~(\ref{eq:explsolChi12}). This
factor leads to the vanishing of functions $\chi_{3,4}$ for
sufficiently long times.\\
To better understand the solutions (\ref{eq:explsolChi12}) and
(\ref{eq:explsolChi34}) we rewrite them in terms of the complex
variable $\beta$
\begin{eqnarray}\label{eq:explsolChi12beta}
\chi_{1,2}(\beta,t)  &=& \frac{1}{2}
\exp\left\{-\frac{|\beta|^{2}}{2} \pm
[\beta\alpha^*(t)-\beta^*\alpha(t)] \right\} , \nonumber \\
\\\label{eq:explsolChi34beta} \chi_{3,4}(\beta,t)  &=& \frac{f(t)}{2}
 \exp\left\{-\frac{|\beta|^{2}}{2} \mp
[\beta\alpha^*(t)+\beta^*\alpha(t)] \right\} , \nonumber \\
\end{eqnarray}
where we have introduced the complex time-dependent function
$\alpha(t)=i\frac{g_{\rm eff}}{\kappa}(1-e^{-\kappa t/2})$. We
immediately recognize that the operators $\rho_{iF}(t)$
corresponding to the functions $\chi_{i}(\beta,t)$ are
\begin{eqnarray}\label{roiFt}
\rho_{1F}(t)& = &\frac{1}{2}|\alpha(t)\rangle\langle\alpha(t)|,
\nonumber
\\ \rho_{2F}(t) & = & \frac{1}{2}|-\alpha(t)\rangle\langle-\alpha(t)|,
\nonumber \\
\rho_{3F}(t)& = &\frac{1}{2}\frac{f(t)}{e^{-2|\alpha(t)|^2}}|
\alpha(t)\rangle\langle-\alpha(t)|,
\nonumber \\
\hspace{1cm}\rho_{4F}(t) & = &
\frac{1}{2}\frac{f(t)}{e^{-2|\alpha(t)|^2}}|-\alpha(t)\rangle
\langle\alpha(t)|.
\end{eqnarray}
\indent We describe now the generation of Schr\"odinger cat states
for the whole atom-field system.  Actually, in the limit of
$\kappa t \ll 1$, when the unitary dynamics dominates over the
incoherent cavity dissipation, $f(t)\simeq e^{-2|\alpha(t)|^2} $
so that the state of the atom-field system is well approximated
by:
\begin{equation}\label{eq:atfsupcat}
|\Psi (t)\rangle_{AF} =
\frac{1}{\sqrt{2}}(|+\rangle|\tilde{\alpha}(t)\rangle +
|-\rangle|-\tilde{\alpha}(t)\rangle),
\end{equation}
with $\tilde{\alpha}(t) = i\frac{g_{\rm eff} t}{2}$.\\
On the other hand, the steady state of the atom-field system is the
mixed state
\begin{eqnarray}\label{eq:roat}
\rho_{AF}^{ss} &=& |+\rangle\langle +|\rho_{1F}^{ss} +
|-\rangle\langle -|\rho_{2F}^{ss} \nonumber \\ &=&
\frac{|+\rangle\langle + ||\alpha^{ss}\rangle\langle\alpha^{ss}| +
|-\rangle\langle -||-\alpha^{ss}\rangle\langle-\alpha^{ss}|}{2} ,
\nonumber \\
\end{eqnarray}
with $\alpha^{ss} = i g_{\rm eff} / \kappa$.

\section{CAVITY FIELD AND ATOM SUBSYSTEM DYNAMICS} 
\begin{figure}[h]
\begin{center}
\includegraphics[scale=0.40]{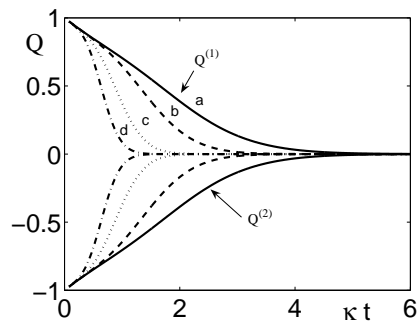}\vspace*{-0.5cm}
\end{center}
\caption{The Mandel-Fano parameter Q for the cavity field versus
dimensionless interaction time in the case of atom detected in the
lower ($Q^{(1)}(t)$) and upper ($Q^{(2)}(t)$) states.  We consider
different values of the mean steady-state photon number $\langle
\hat{N}\rangle_{SS}$: (a) 1, (b) 2, (c) 5, (d) 10. \vspace*{1cm}}
\label{fig:2Mandel}
\end{figure}

We consider the
reduced density operator for the cavity field $\rho_{F}(t) =
\textrm{Tr}_{A}\lbrack \rho_{AF}(t) \rbrack = \rho_{1F}(t) +
\rho_{2F}(t)$, where $Tr_A$ denotes the partial trace over the
atomic variables. From Eq.~(\ref{roiFt}) we obtain
\begin{equation}\label{eq:lascavfield}
\rho_{F}(t)
=\frac{|\alpha(t)\rangle\langle\alpha(t)|+|-\alpha(t)\rangle\langle
-\alpha(t)|}{2} \, .
\end{equation}
and we see that it is always a mixed state. The cavity field mean
photon number after an interaction time $t$ is
\begin{equation}
\label{meanNth}\langle \hat{N}\rangle(t) = Tr_F \lbrack
\hat{a}^{\dagger}\hat{a}\rho_F(t) \rbrack = |\alpha(t)|^2=
\frac{g_{\rm eff}^2}{\kappa^2}(1-e^{-\kappa t/2})^2.
\end{equation}
In the steady state the cavity field mean photon number is given by
$\langle \hat{N}\rangle^{SS}=(g_{\rm eff} / \kappa)^2 $, that is,
the squared ratio between the effective coupling frequency and the
cavity decay rate, which rule the coherent and incoherent regimes of
cavity field dynamics, respectively. The time-dependent photon
number distribution $p_n(t)$ is given by a Poissonian distribution
\begin{equation}
\label{pnt} p_n(t)=\frac{|\alpha(t)|^{2n}}{n!} e^{-|\alpha(t)|^2}.
\end{equation}
Hence, at any time the photon number distribution of the SDOAL is
that of a coherent field, a natural consequence of tracing
orthogonal atomic states $| \pm \rangle$. Certainly, this will not
be the case if we make a projective atomic measurement in the bare
basis $\{ | 1\rangle , | 2 \rangle \}$ at a time $t$ during the
transient. Actually, after the atom measurement, the cavity field is
in either of the pure states
\begin{equation}
\label{ROF12}
\rho_F^{(1,2)}(t)=\frac{\rho_{1F}(t)+\rho_{2F}(t)\pm[\rho_{3F}(t)+\rho_{4F}(t)]}{2p_{1,2}(t)}
\end{equation}
where $p_{1,2}(t)$ is the probability to find the atom in the state
$|1\rangle$, $|2\rangle$ respectively at a time t (see below). The
corresponding photonstatistics are
\begin{eqnarray}\label{eq:pn12}
p_n^{(1,2)}(t) &=& \langle n|\rho_F^{(1,2)}(t) |n\rangle \nonumber \\
&=& \frac{1}{1\pm f(t)} e^{-|\alpha(t)|^2} \frac{|\alpha(t)|^{2n}}{n!}\times \nonumber \\
&\times &\left[1\pm(-1)^n\frac{f(t)}{e^{-2|\alpha(t)|^2}} \right]
\end{eqnarray}
In the transient dynamics, for times $kt\ll 1$,
$\alpha(t)=\tilde{\alpha}(t)$ and the cavity field states are even
and odd cat states
\begin{equation}
\label{cat12}
|\psi(t)\rangle_F^{(1,2)}=\frac{|\tilde{\alpha}(t)\rangle \pm
|-\tilde{\alpha}(t)\rangle}{\sqrt{2(1\pm e^{-2|\alpha(t)|^2})}}
\end{equation}
As is well known~\cite{gerry}, states as in Eq.~(\ref{cat12}) can
exhibit quantum effects including oscillating photonstatistics, sub
Poissonian photonstatistics, and quadrature squeezing. In
Fig.~\ref{fig:2Mandel} we show the time behaviour of the Mandel-Fano
parameter $Q=\frac{\langle \hat{N}^2 \rangle -\langle \hat{N}
\rangle^2}{\langle \hat{N} \rangle}-1$ in both cases of atom
detected in the lower ($Q^{(1)}(t)$) and upper ($Q^{(2)}(t)$) state
and for different values of the steady state mean photon number. We
see that $Q^{(1)}(t)$ and $Q^{(2)}(t)$ exhibit super and sub
Poissonian photonstatistics, respectively, before approaching the
steady state Poissonian distribution.
\\
\indent Now, we consider the reduced density operator for the atom
$\rho_{A}(t)=Tr_F \lbrack \rho_{AF}(t) \rbrack$. From
Eq.~(\ref{roiFt}) we derive the following density matrix in the
basis $\{|+\rangle,|-\rangle\}$
\begin{equation}
\label{rhoA} \rho_{A}^{\pm}(t)=\frac{1}{2}\left(\begin{array}{cc} 1
&
f(t)\\
f(t) & 1
\end{array}\right).
\end{equation}
From this atomic density matrix we can derive the probabilities
$p_{1,2}$ to find the atom in the lower or upper state,
\begin{equation}
\label{p1} p_{1,2}(t)=\langle 1,2|\rho_ {A}(t) |1,2\rangle=
\frac{1}{2}[1\pm f(t)].
\end{equation}
We observe that in the steady state the atomic population of the
upper level $|3\rangle$ is zero and those of the lower and
intermediate levels are both equal to 0.5. The physical intuition
behind this result is the orthogonality of coherent states $| \alpha
(t) \rangle$ and $|- \alpha (t) \rangle$ when $t \rightarrow
\infty$.
We will employ these results in the following section to study the
entanglement properties and the decoherence of the system.

\section{ ENTANGLEMENT AND DECOHERENCE ANALYSIS}
\begin{figure}[h]
\begin{center}
\includegraphics[scale=0.40]{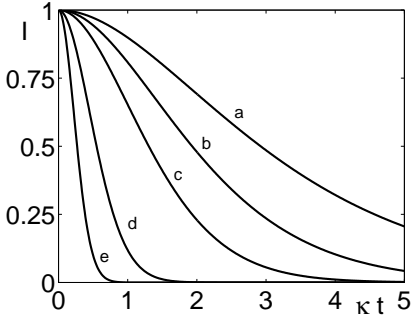}\vspace*{-0.5cm}
\end{center}
\caption{Atomic population inversion $I$ versus dimensionless time
evaluated for different values of $\langle \hat{N}\rangle_{SS}$:
(a) 0.25, (b) 0.5, (c) 1, (d) 5, (e) 20.} \vspace*{-0.5cm}
\label{fig:3decoherence}
\end{figure}

In sections III,  we presented a new scheme for generating
atom-field superposition states [see Eq.~(\ref{eq:atfsupcat})] in
the transient regime and we described the steady state of a SDOAL.
Now, we evaluate atom-field entanglement properties and show how
to monitor the decoherence towards a steady state. We have shown
that the state of the whole atom-field system is almost a pure
state on a time scale much shorter than the cavity decay time
$1/k$. Therefore, in this case, we can use the entropy of
entanglement $E(\Psi)$ as an entanglement measure. It can be
calculated in a straightforward way using the
equality~\cite{intro2},
\begin{equation}
E(\Psi)\equiv S_A=S_F,
\end{equation}
where $S_A$ and $S_F$ denote the von Neumann entropy of the atomic
and field subsystems, respectively. The atomic entropy reads
\begin{equation}
S_A = -\lambda_1 log_2\lambda_1-\lambda_2 log_2\lambda_2,
\end{equation}
where $\lbrace \lambda_1,\lambda_2 \rbrace$ are the eigenvalues of
the reduced atomic density matrix $\rho_A(t)$. In the limit $\kappa
t \ll 1$, the atomic density matrix in Eq.~(\ref{rhoA}) can be
approximated by
\begin{equation}
\label{rhoApure}
\tilde{\rho_{A}}(t)=\frac{1}{2}\left(\begin{array}{cc}
1 & e^{-2|\tilde{\alpha}(t)|^2}\\
e^{-2|\tilde{\alpha}(t)|^2} & 1
\end{array}\right),
\end{equation}
whose eigenvalues are
\begin{equation}
\label{eigens} \lambda_{1,2}(t)=\frac{1}{2}\left[1\pm
e^{-2|\tilde{\alpha}(t)|^2} \right]=\frac{1}{2}\left[1\pm
e^{-\langle \hat{N}\rangle^{SS}(\kappa t)^2/2]} \right].
\end{equation}
\begin{figure}[h]
\begin{center}
\includegraphics[scale=0.40]{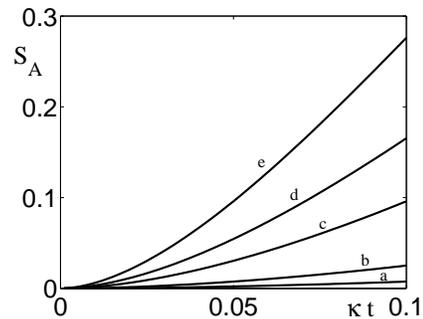}\vspace*{-0.5cm}
\end{center}
\caption{The Von Neumann entropy $S_A$ for the atom-field system
versus dimensionless time for different values of the steady-state
mean photon number $\langle \hat{N}\rangle_{SS}$: (a) 0.25, (b) 1,
(c) 5, (d) 10 , (e) 20.} \vspace*{-0.5cm}
\label{fig:2entanglement}
\end{figure}
In Fig.~\ref{fig:2entanglement} we plot the time evolution of the
von Neumann entropy $S_A$ for different values of the steady-state
mean photon number $\langle \hat{N}\rangle^{SS}$. We see that the
system gets more entangled for larger values of $\langle
\hat{N}\rangle^{SS}$, i.e., when the ratio $g_{\rm eff}/k$ is
large.\\
\indent The next question is how to monitor the decoherence of the
whole atom-field system, that is, the reduction from a pure state
to a statistical mixture. The environment-induced decoherence of a
cavity field prepared in a superposition state has been both
theoretically and experimentally studied in the case of high-Q
microwave cavities~\cite{intro13},~\cite{Davidovich}. We remark
that the cavity field reduced density operator does not depend on
the decoherence function $f(t)$. However, a simple way to monitor
the decoherence of the atom-field system is to measure the atomic
populations $p_{1,2}(t)$ of Eq.~(\ref{p1}). In fact, the atomic
inversion $I(t)=p_{1}(t)-p_{2}(t)$ is exactly the function $f(t)$
(Eq.~(\ref{eq:ft})), that can be rewritten as:
\begin{equation}
\label{ftmn} f(t)=\exp[-2\langle \hat{N} \rangle^{SS} \kappa t +
4\langle \hat{N} \rangle^{SS} (1-e^{-\kappa t / 2})].
\end{equation}
In Fig.~\ref{fig:3decoherence} we illustrate the time evolution of
the atomic inversion $I(t)$ for different values of $\langle
\hat{N}\rangle^{SS}$ showing that the decoherence dynamics is rather complex.\\
For dimensionless times $\kappa t \ll 1$ the inversion $I(t)$
shows a Gaussian fall-off as $\exp\{-\langle \hat{N}\rangle^{SS}
\kappa ^2 t^2/2 \}$, independent of the cavity field decay rate
$k$. We recall that in this limit the interaction generates the
atom-field cat-like superposition as in Eq.~(\ref{eq:atfsupcat}).
After this transient, the effective decoherence process begins in
correspondence to the inflection point at time:
\begin{equation}\label{inflect}
t_F =-\frac{2}{\kappa} \ln (1+\frac{1-\sqrt{1+16\langle
\hat{N}\rangle^{SS}}}{8\langle \hat{N}\rangle^{SS}} ).
\end{equation}
For $\langle \hat{N} \rangle^{SS}\gg 1$, corresponding to
effective strong coupling conditions, we have $t_F\cong
\frac{1}{\kappa\sqrt{\langle \hat{N} \rangle^{SS}}}=g_{\rm
eff}^{-1}$, and the decoherence function $f(t)$ can be well
approximated by:
\begin{equation}\label{expfit}
I(t_F)\exp\{-2\kappa \langle \hat{N}\rangle^{SS}(1-e^{-\kappa
t_F/2})(t-t_F)\}.
\end{equation}
Hence, we can introduce the decoherence rate
\begin{eqnarray}\label{decayrate}
\gamma_D &=& \kappa \frac{\sqrt{1+16\langle \hat{N}
\rangle^{SS}}-1}{4}\nonumber\\
&\cong& \kappa \sqrt{\langle \hat{N} \rangle^{SS}}=g_{\rm eff}.
\end{eqnarray}
We note that the decoherence rate is given by the effective
coupling constant and hence it is independent of cavity
dissipation.\\
\indent In the opposite limit of small $\langle
\hat{N}\rangle^{SS}$ we have $t_F \gg k^{-1}$; the system is close
to the mixed steady state and we recover an exponential decay with
a decoherence rate $\gamma'_D = 2 \kappa \langle \hat{N}
\rangle^{SS}$. This is the standard decoherence rate for cat
states of the cavity field alone~\cite{Davidovich}, which can be
generated in our system after an atomic measurement.

\section{NUMERICAL SIMULATIONS}
\begin{figure}[h]
\begin{center}
a)\includegraphics[scale=0.40]{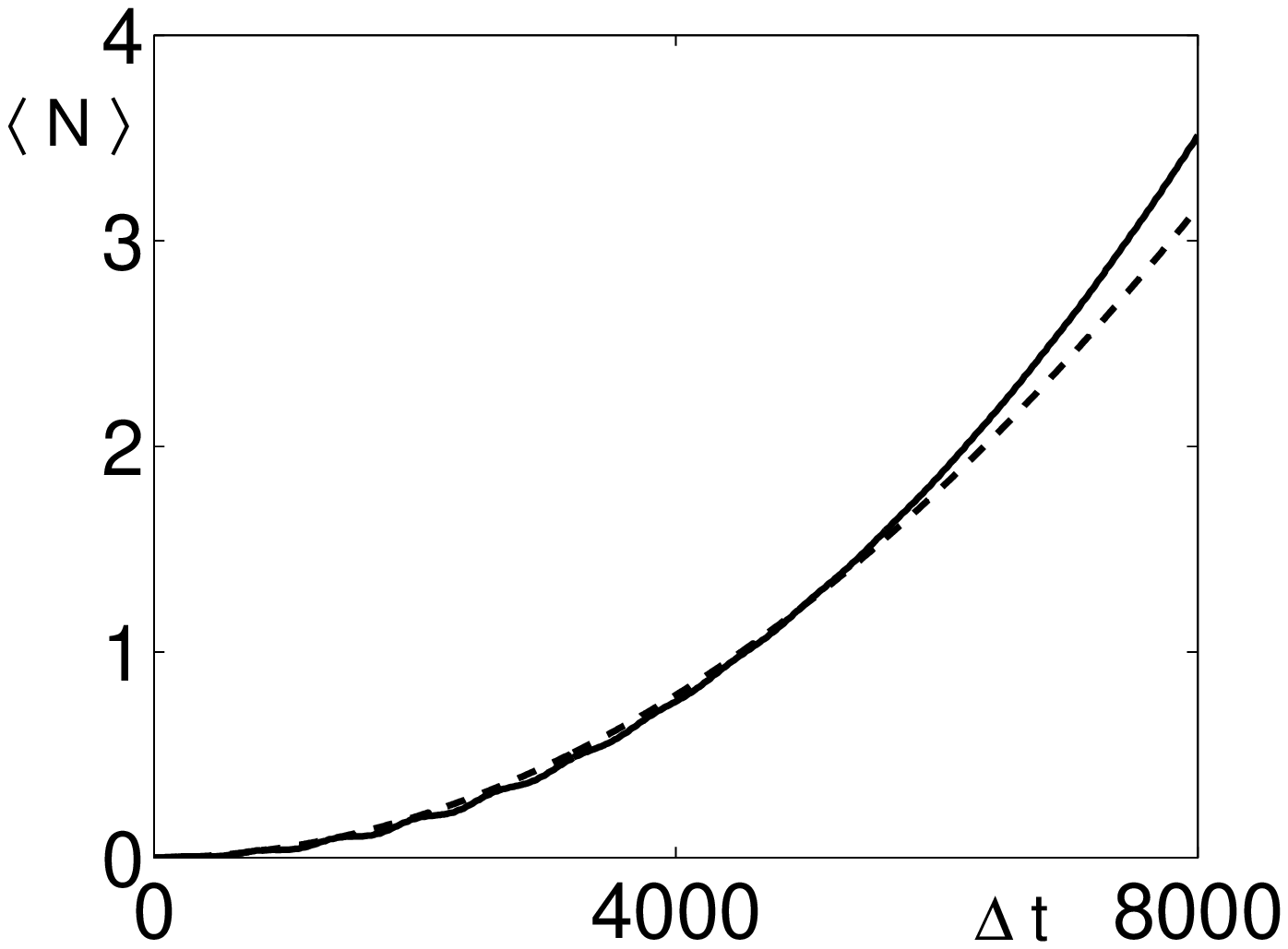}\\
b)\includegraphics[scale=0.40]{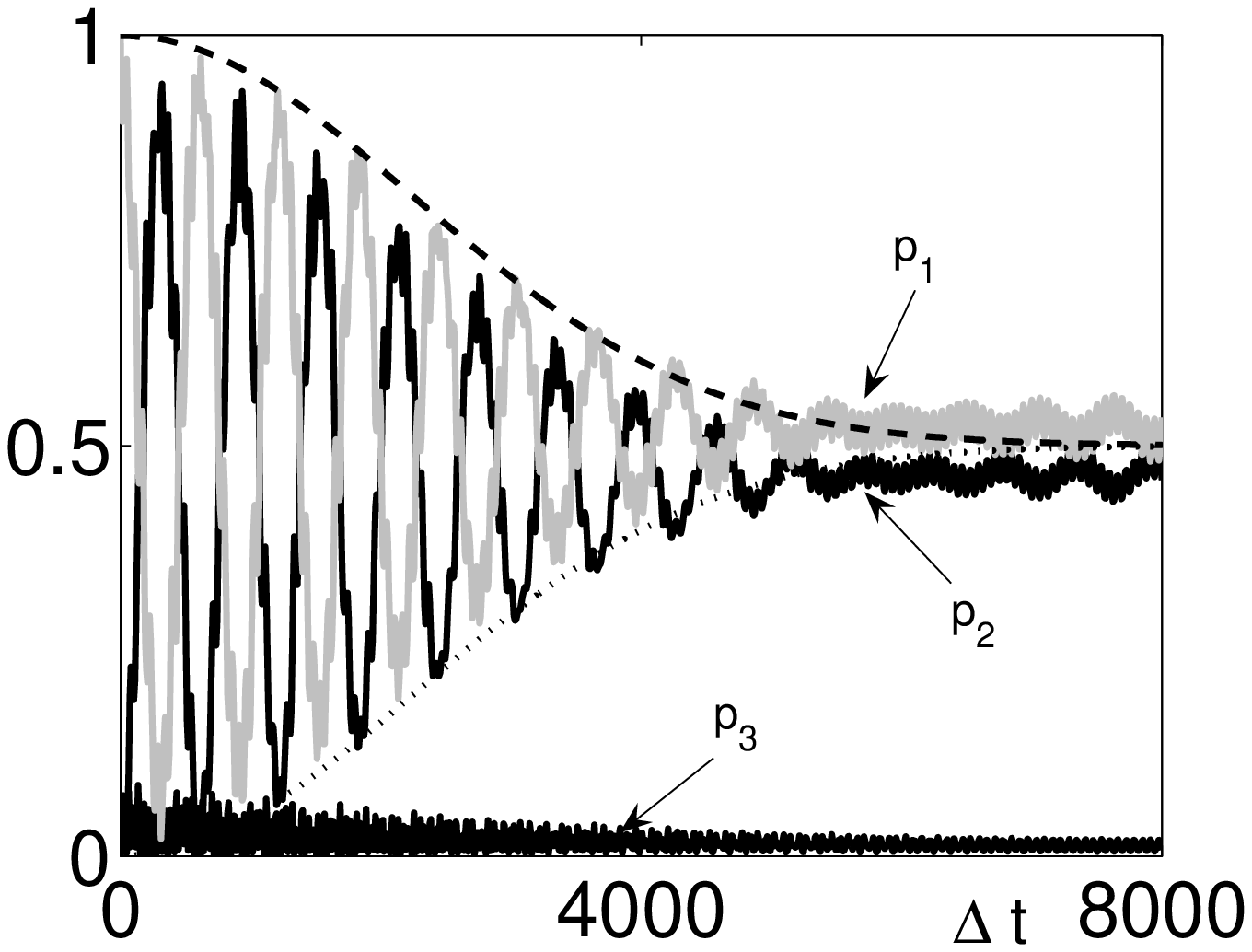}\vspace*{-0.5cm}
\end{center}
\caption{Hamiltonian dynamics of the full three-level system for the
parameters $\tilde{\Delta'}=0.9$, $\tilde{g}=0.004$,
$\tilde{\Omega}=0.1$, $\tilde{\Omega'_1}=0.05$,
$\tilde{\Omega'_2}=0.1$.  a) Cavity field mean photon number vs
dimensionless time: numerical value (solid line) and theoretical
value (dashed line). b) Atomic populations: numerical values (solid
lines) and theoretical values (dashed lines).} \vspace*{-0.5cm}.
\label{fig:5Hamiltonian}
\end{figure}

\begin{figure}[h!]
\begin{center}
\includegraphics[scale=0.40]{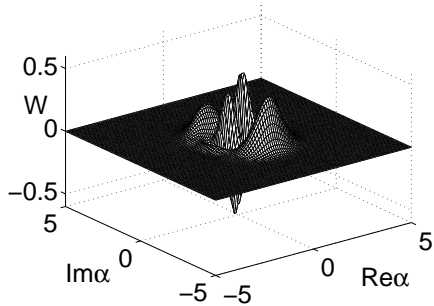}\\
\end{center}
\caption{Wigner function for the cavity field state after detection
of the atom in the ground state $|1\rangle$. The parameters are as
in Fig.~\ref{fig:5Hamiltonian} and the dimensionless time is $\Delta
t=7160$.} \vspace*{-0.5cm}. \label{fig:6Felix}
\end{figure}

\begin{figure}[h]
\begin{center}
a)\includegraphics[scale=0.40]{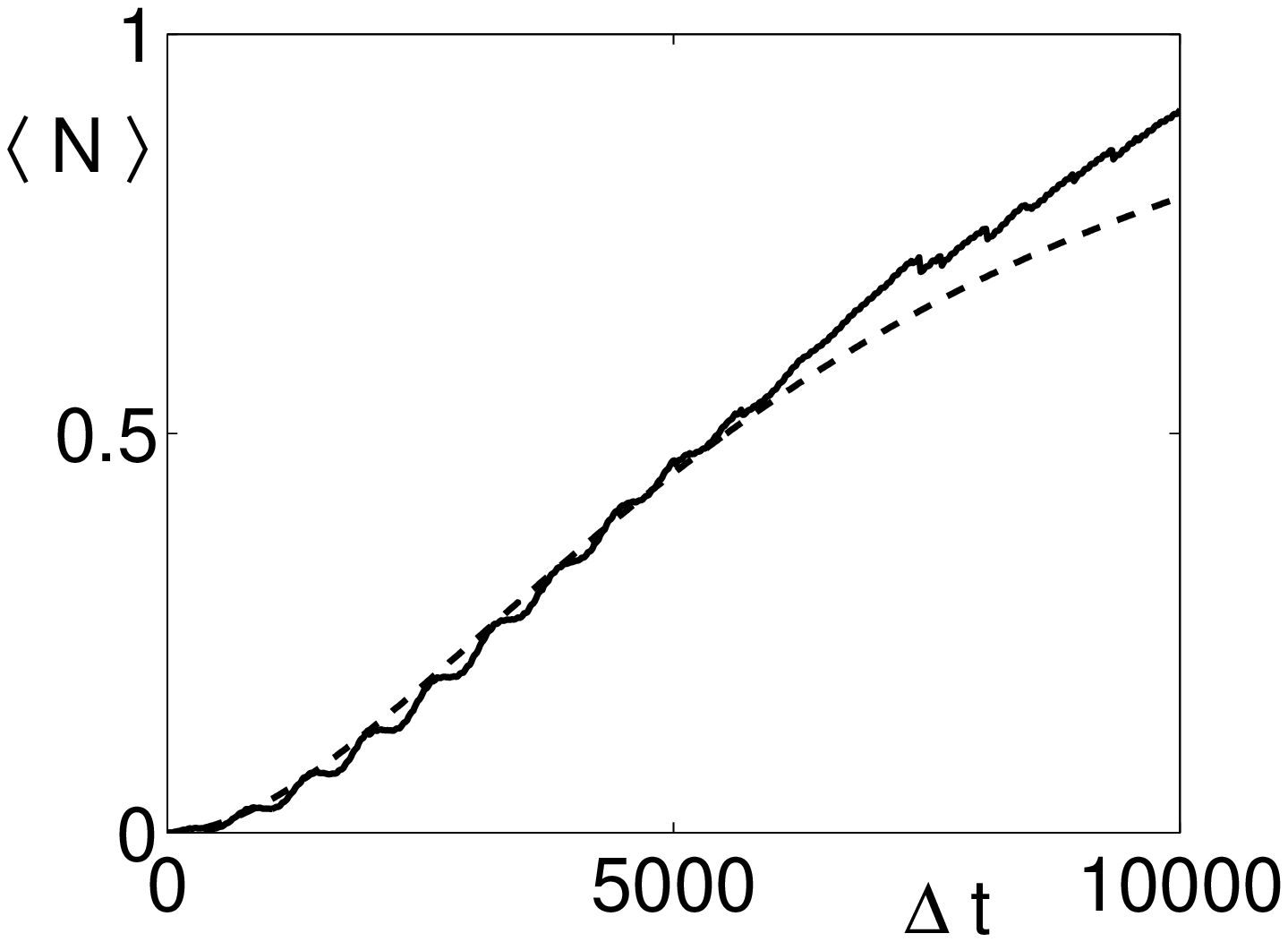}\\
b)\includegraphics[scale=0.40]{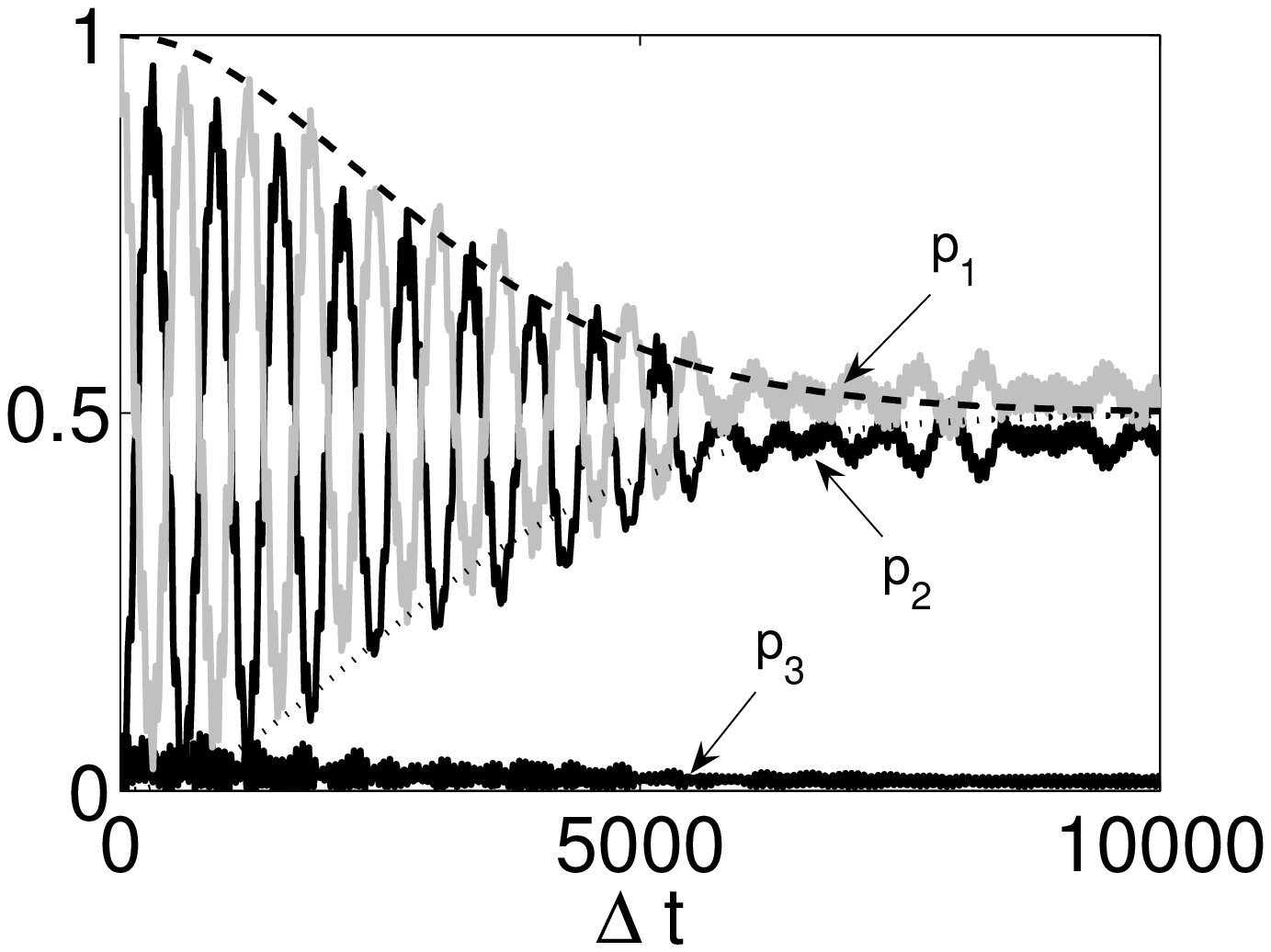}\vspace*{-0.5cm}
\end{center}
\caption{Full dynamics of the three-level system for the parameters
as in Fig.~\ref{fig:5Hamiltonian}, and for $\tilde{k}=\tilde{g}_{\rm
eff}=0.00044$. a) Cavity field mean photon number vs dimensionless
time: numerical value (solid line) and theoretical value (dashed
line). b) Atomic populations: numerical values (solid lines) and
theoretical values (dashed lines). We used twenty trajectories.}
\vspace*{-0.5cm}. \label{fig:7dissipative}
\end{figure}

In this section we discuss the theoretical approach presented in
the above sections from a numerical point of view. In fact, by
means of a first order perturbative approach we have reduced the
full three-level system dynamics to an effective two-level one
described by the ME in Eq.~(\ref{eq:maseqL}). We now discuss the
validity of that approximation for both Hamiltonian and
dissipative dynamics.\\
\indent In the numerical analysis we need to solve the full system
ME
\begin{equation}
\label{eq:maseqNUM} \dot{\rho}_{AF}  =  -
\frac{i}{\hbar}[\hat{\mathcal{H}}_{i}(t), \rho_{AF}] +
\hat{\mathcal{L}}\rho_{AF},
\end{equation}
where the system Hamiltonian is given in
Eq.~(\ref{H_interactionRWA}) and the dissipative process is ruled
by the Liouville super operator in Eq.(\ref{eq:dissip}). To solve
Eq.~(\ref{eq:maseqNUM}) numerically we consider the dimensionless
time $\tilde{t}=\Delta t$ and the following dimensionless parameters
\begin{eqnarray}
\label{dimparNUM} \tilde{\Delta}' &=&
\frac{\Delta'}{\Delta},\hspace{0.3cm}\tilde{g} =
\frac{g}{\Delta},\hspace{0.3cm}\tilde{\Omega} =
\frac{\Omega}{\Delta},\hspace{0.3cm}\tilde{\Omega'_1} =
\frac{\Omega'_1}{\Delta}\hspace{0.3cm}\tilde{\Omega'_2} =
\frac{\Omega'_2}{\Delta}\nonumber\\
\tilde{\kappa} &=&
\frac{\kappa}{\Delta}, \hspace{0.3cm}\tilde{g}_{\rm eff}=\frac{\tilde{g}
\tilde{\Omega}}{\tilde{\Delta}'},\hspace{0.3cm}\tilde{\Omega}_{\rm
eff}=\frac{\tilde{\Omega}'_2 \tilde{\Omega}}{\tilde{\Delta}'} .
\end{eqnarray}
In order to solve the ME by means of the Monte Carlo Wave Function approach
(MCWF)~\cite{Dalibard}, we rewrite the ME in the
Lindblad form to identify the collapse and the "free evolution"
operators
\begin{eqnarray}
\label{eq:maseqLFNUM}\dot{\rho}_{AF} & = & -
\frac{i}{\hbar}(\hat{\mathcal{H}}_{e}
\rho_{AF}-\rho_{AF}\hat{\mathcal{H}}_{e}^{\dagger}) +
\hat{C}\rho_{AF}\hat{C}^{\dagger}
\end{eqnarray}
where the non-Hermitian effective Hamiltonian
$\hat{\mathcal{H}}_{e}$ is given by
\begin{equation}
\label{NorHermitianHNUM}\hat{\mathcal{H}}_{e}  =
\frac{\hat{\mathcal{H}}_{i}(\tilde{t})}{\Delta} -
\frac{i\hbar}{2}\hat{C}^{\dagger}\hat{C},
\end{equation}
and the only one collapse operator is
$\hat{C}=\sqrt{\tilde{k}}\hat{a}$.
The system dynamics can be
simulated by a suitable number of trajectories, i.e. stochastic
evolutions of the wave function $|\psi(\tilde{t})\rangle$, by means
of the following main rule
\begin{equation}
\label{MCWF} |\psi(\tilde{t}+\delta
\tilde{t})\rangle=\big\{\begin{array}{c}
\frac{(1-\frac{i}{\hbar}\hat{\mathcal{H}}_{e}\delta
\tilde{t})|\psi(\tilde{t})
\rangle}{\sqrt{1-\delta_p(\tilde{t})}}\hspace{1cm}
if \hspace{0.2cm} \delta_p(\tilde{t})<N_{rnd} \\
\frac{\hat{C}_i|\psi(\tilde{t})\rangle}{\sqrt{\delta_{p}
(\tilde{t})}}\hspace{1cm}
if \hspace{0.2cm} \delta_p(\tilde{t})> N_{rnd} \\
\end{array},
\end{equation}
where $\delta\tilde{t}$ is a suitable small time interval,
$\delta_p(\tilde{t})$ is the collapse probability at time
$\tilde{t}$, and $N_{rnd}$ is a random number generated from a
uniform distribution in $[0,1]$. We note that the collapse
probability depends on the cavity field mean photon number
$\langle\hat{N}\rangle(\tilde{t})$ and can be evaluated as
$\delta_{p}(\tilde{t})=\delta\tilde{t}\tilde{k} \langle
\hat{N}\rangle(\tilde{t})$. In the simulations we must consider
parameters values in agreement with the theoretical conditions
required by the two small rotations.\\
\indent First we discuss the full three-level system Hamiltonian
dynamics ($k=0$) in order to confirm the validity of the effective
two-level Hamiltonian of Eq.~(\ref{eq:keyLas}). We consider the time
evolution of the cavity field mean photon number and of the atomic
populations, and we compare the numerical results with the
theoretical expressions $\langle\hat{N}\rangle(t)=\frac{g_{\rm
eff}^2 t^2}{4}$, and $p_{1,2}(t)=\frac{1}{2}[1\pm \exp(-\frac{g_{\rm
eff}^2 t^2}{2})]$, $p_{3}(t)=0$. As an example, we show in
Fig.~\ref{fig:5Hamiltonian} a case where the ratio of the effective
parameters is $\frac{\Omega_{\rm eff}}{g_{\rm eff}}=25$. We see a
good agreement for the mean photon number
(Fig.~\ref{fig:5Hamiltonian}a). The theoretical functions
$p_{1,2}(t)$ fit the envelopes of the numerical fast oscillating
populations (Fig.~\ref{fig:5Hamiltonian}b). In fact, in the
numerical analysis we do not take into account the RWA
approximation. In particular, the populations of levels $|1\rangle$
and $|2\rangle$ approach the expected value of $0.5$, and the
population of the upper level $|3\rangle$ is always negligible.
In addition, we tested the prediction that the effective dynamics
allows to generate cavity field cat states when the atom is measured
in level $|1\rangle$ or $|2\rangle$. In Fig.~\ref{fig:6Felix} we
show the Wigner function that describes in phase space the cavity
field state prepared by an atomic measurement in level $|1\rangle$
and we see the typical features of a cat state.\\
\indent Now we consider the full dynamics including dissipation of
Eq.~(\ref{eq:maseqNUM}), for the same parameters as in
Figs.~\ref{fig:5Hamiltonian}, \ref{fig:6Felix}, and with
$\tilde{k}=\tilde{g}_{\rm eff}$, so that we expect that the steady
state value of mean photon number is one and it is reached in a
time that is twice that of the atomic populations. In
Fig.\ref{fig:7dissipative}a we compare the numerical results for
the time evolution of the cavity field mean photon number to the
theoretical behavior predicted by Eq.~(\ref{meanNth}), showing
that there is a good agreement. In Fig.~\ref{fig:7dissipative}b we
consider the numerically simulated time evolution of the atomic
populations $p_{j}(t)$ ($j=1,2,3$) compared to the theoretical
functions in Eq.~(\ref{p1}).
We remark that the population of the upper level $p_{3}(t)$ always
remains negligible, the populations $p_{1}(t)$ and $p_{2}(t)$ reach
the steady state value of 0.5, and the theoretical curves fit the
envelopes of the fast oscillating functions. The above results
provide a clear demonstration of the validity of the two-level
approximation developed in section II, which is at the basis of the
subsequent theoretical developments.

\section{CONCLUSIONS}
We have introduced a solvable model of a strongly-driven one-atom
laser in the optical regime of cavity QED. We have shown
analytically and numerically that the complex dynamics of a
three-level atom, dispersively coupled to an optical cavity mode
and to three laser fields, can be well approximated by a two-level
atom that is resonantly coupled to a cavity mode and a strong
coherent field. The effective coupling constant is a combination
of the atom-cavity field coupling constant, the amplitude of one
of the external lasers and the detuning parameter. The initial
transient regime shows that the system is approximately in an
entangled atom-cavity field state, a Schr\"{o}dinger cat state,
and we show that the amount of entanglement depends on the steady
state mean photon number that is the ratio between the effective
coupling constant and the cavity decay rate. In addition, we
propose a scheme for monitoring the whole system decoherence based
on atomic population measurements. We find that, for large values
of the steady-state mean photon number (i.e. in the strong
coupling regime), and for time larger that the inverse of the
effective coupling constant, the decoherence behavior can be well
approximated by an exponential decay whose rate is given by
the effective coupling constant.\\
\indent The cavity field subsystem is always in a mixed state
whose photon number distribution is Poissonian, while the atomic
subsystem can exhibit coherences. If we measure the atomic state
at a given time, we can project the cavity field in a cat-state
with sub-Poissonian or super-Poissonian photon statistics
depending on the detected atomic state.

\section{Acknowledgments}
PL acknowledges support from The Hearne Institute for Theoretical
Physics, The National Security Agency, The Army Research Office and
The Disruptive Technologies Office. ES acknowledges support from EU
EuroSQIP and DFG SFB 631 projects. AL thanks the support from MIUR
through the project PRIN-2005024254-002.

\end{document}